# A lossless *a priori* splitting rule for split-delivery routing problems

Bo Jones[1], Julien Yu[1], and John Gunnar Carlsson[*1]

[1]Epstein Department of Industrial and Systems Engineering, University of Southern California

**Abstract**

Resource allocation problems in which demand is splittable are usually solved using different solution methods from their unsplittable equivalents. Although splittable problem instances can be the easier of the two (for example, they might simply correspond to a linear relaxation of a discrete problem), there exist many problems, including routing problems, for which the converse is true: that is, the technology for solving unsplittable problems is mature, but the splittable counterpart is not. For such problems, one strategy that has recently shown potential is the use of an *a priori splitting rule* in which each customer's demand is split into smaller pieces in advance, which enables one to simply solve the splittable problem as an instance of the unsplittable version. An important factor to consider is the number of pieces that result after this splitting: a large numbers of pieces will allow more splitting patterns to be realizable, but will result in a larger problem instance. In this paper, we introduce a splitting rule that minimizes the number of pieces, subject to the constraint that all demand splitting patterns remain feasible. Computational experiments on benchmark instances for the vehicle routing problem and a time-windows extension show that the solution quality of our proposed splitting rule can match the performance of existing approaches.

## 1 Introduction

A recurring theme in resource allocation problems is the distinction between problems having splittable versus unsplittable demand. For many problems, the splittable formulation is easier because it is merely a linear programming relaxation of the unsplittable formulation; for example, the maximum flow problem becomes NP-hard when splitting is forbidden [24].

Within the realm of vehicle routing, introducing the option to split demand generally *increases* the difficulty of the routing problem [9], although it can obviously improve the quality of solutions [8]. As a consequence, splittability is not a supported option in many off-the-shelf VRP solvers, even those with otherwise sophisticated feature sets [21, 23, 28, 31]. In order to address this, the paper [17] introduces a transformation of the split-delivery capacitated vehicle routing problem to an unsplittable VRP, which they call an *a priori splitting rule*. The reduction is simple: divide each demand node into multiple demand nodes at the same location, with the sum of the demands of these copies being equal to the original demand at the location. Then, solve an unsplittable CVRP on this larger problem.

[*]Corresponding author, jcarlsso@usc.edu.

When one employs an a priori splitting rule, a natural trade-off quickly presents itself: at one extreme, if a demand node is split into a small number of pieces, the number of possible ways to split demand between the vehicles is limited. For example, if we split a node with demand of 10 into three nodes with demand 1, 3, and 6, but the true (unknown) optimal solution to the problem requires that that node be split between two vehicles that service 2 and 8 units of demand respectively, then we are out of luck, and the rule returns a sub-optimal solution. At the other extreme, if a demand node is split into a large number of pieces (for example, splitting a node with a demand of 10 into 10 nodes with demand 1), then the set of allowable solutions is obviously larger, at a cost of greater problem complexity. The paper [17] introduces an elegant heuristic splitting rule that manages the overall problem size, while maintaining a wide range of solutions. This is accomplished by splitting nodes into a mixture of small and large pieces that is motivated by US currency denominations (quarters, dimes, nickels, and pennies).

This paper presents what we call a *lossless* a priori splitting rule that results in the smallest possible computational overhead, which is to say, it splits each demand node into the minimum number of pieces. By "lossless", we mean that the true (unknown) optimal solution is never lost, because all (integral) allocations of demand to vehicles remain feasible. In this way our goal is philosophically similar to the problem of sparse signal recovery that motivates work in $\ell_1$ approximation of $\ell_0$ minimization [15, 34]. We want to use a small number of components to capture some unknown original thing with high fidelity. In our case we find the smallest number of components such that we in fact lose nothing.

Our method is applicable to any other problem for which splitting has benefit and the split-delivery version is harder than the non-split version. These include arc routing [26], facility location problems [12], and bin packing [14, 16]. However, our focus will be on routing problems. We will apply our lossless splitting rule to the Split-Delivery Capacitated Vehicle Routing Problem and the Split-Delivery Capacitated Vehicle Routing Problem with Time Windows.

This paper is structured as follows: Section 2 presents the heuristic solution of [17] that inspired this paper, Section 3 presents a lossless and optimal *a priori* demand splitting rule, and Section 4 compares the performance of our rule to algorithms designed for the split-delivery version of our routing problems.

## 2 Prior work and preliminaries

The problem that motivated the introduction of the a priori split rule is an extension of the *Capacitated Vehicle Routing Problem:*

**Definition 1** (Split-Delivery Vehicle Routing Problem)**.** The problem instance takes the form of a directed graph $G = (N, A)$ where $N = \{N_0, \ldots, N_n\}$ is a set of nodes representing a depot, $N_0$, and $n$ customers and $A = \{(N_i, N_j) : N_i, N_j \in N, i \neq j\}$ is the set of all arcs between customers. Associated with each customer node $N_i$ is a demand $D_i$. Associated with each arc $(N_i, N_j)$ is a travel cost $c_{ij}$. There are $k$ trucks that are are known to be sufficient to meet all demand. Each truck has capacity $Q$. Trucks will take tours on $G$ that begin and end at the depot, filling all customer demands and incurring a travel cost, the sum of the traversed arcs' costs, which we would like to minimize. Customer demands can be split across trucks, with multiple trucks then making a visit to a given customer.

The paper [17] gives two a priori rules for splitting demand nodes. The first rule is called the *20/10/5/1 rule*. If each vehicle has capacity $Q$, their rule splits each demand node $D_i$ into $m_{20}$ pieces of $0.20Q$, $m_{10}$ pieces of $0.10Q$, $m_5$ pieces of $0.05Q$, $m_1$ pieces of $0.01Q$ and at most one remaining piece of size less than $0.01Q$, by letting

$$\begin{aligned} m_{20} &= \max\{m \in \mathbb{N} \cup \{0\} : 0.20Qm \leq D_i\}, \\ m_{10} &= \max\{m \in \mathbb{N} \cup \{0\} : 0.10Qm \leq D_i - 0.20Qm_{20}\}, \\ m_5 &= \max\{m \in \mathbb{N} \cup \{0\} : 0.05Qm \leq D_i - 0.20Qm_{20} - 0.10Qm_{10}\}, \\ m_1 &= \max\{m \in \mathbb{N} \cup \{0\} : 0.01Qm \leq D_i - 0.20Qm_{20} - 0.10Qm_{10} - 0.05Qm_5\}. \end{aligned}$$

As an example, if vehicles have capacity of 200, then a node with demand $D_i = 157$ will be be divided into $m_{20} = 3$ nodes of demand $0.20Q = 40$, $m_{10} = 1$ node of demand $0.1Q = 20$, $m_5 = 1$ node of demand $0.05Q = 10$, and $m_1 = 1$ node of demand 7. The other rule is the *25/10/5/1 rule*, for which the construction is analogous.

The authors test their heuristic solutions against those of [5, 6, 10, 11, 18, 19, 32] on a number of the same SDVRP benchmark instances. In their analysis, they find that "for the 82 benchmark instances, it is much faster than the state-of-the-art algorithms and often produces solutions with comparable quality." In fact, in a handful of cases it hit upon the best known solution in a reasonable amount of time. It is this success that motivates our present analysis.

Most of the current best known solutions for SDVRP benchmark instances were found by the algorithm proposed by [32], who implemented a multi-start Iterated Local Search (ILS) based heuristic with special perturbation mechanisms. They outperformed the BKS of previous approaches [5, 6, 19] for 243 out of 324 instances, by an astounding average of 1.15%. Though [17] also found a few BKS, they still have an average gap of 0.74%, 2.24% and 1.50% when compared to [32] on the three sets of instances they tested.

We also apply our a priori split rule to an extension of the *Split-Delivery Vehicle Routing Problem:*

**Definition 2** (Split-Delivery Vehicle Routing Problem with Time Windows)**.** The problem instance takes the form of a directed graph $G = (N, A)$ where $N = \{N_0, \ldots, N_n\}$ is a set of nodes representing a depot $N_0$ and $n$ customers, and $A = \{(N_i, N_j) : N_i, N_j \in N, i \neq j\}$ is the set of all arcs between customers. Associated with each customer node $N_i$ is a demand $D_i$, a time window $[a_i,\ b_i]$, and a service time $s_i$. Associated with each arc $(N_i, N_j)$ is a travel cost $c_{ij}$ and a travel time $t_{ij}$. There are $k$ trucks that are are known to be sufficient to meet all demand. Each truck has capacity $Q$. Trucks will take tours on $G$ that begin and end at the depot, filling all customer demands and incurring a travel cost, the sum of the traversed arcs' costs, which we would like to minimize. Customer demands can be split across trucks, with multiple trucks then making a visit to a given customer. All trucks visiting a customer must arrive at the customer within the customer's time window and remain at the customer for a service time.

The current best known solutions for SDVRPTW benchmark instances were obtained by [7, 20, 25]. Desaulniers [20] uses an exact branch-and-price-and-cut method. Archetti et al. [7] introduced enhancement procedures for this algorithm. Luo et al. [25] developed a branch-and-price-and-cut algorithm for the problem along with an extension to weight-dependent travel costs. A similar problem, but one in which demand is split according to predetermined discrete orders, is solved by Salani and Vacca [30].

## 3 Minimum size coalescing partitions

Recall that an *integer partition* of a positive integer $n$ is a way of writing $n$ as a sum of unordered positive integers. For example, a valid partition of $n = 11$ is $\{4, 3, 2, 2\}$. A partition that consists of $k$ elements is called a *$k$-partition.* We adopt the standard notational convention from combinatorics that $\mu \vdash n$ if $\mu$ is a partition of $n$, that is, $\sum_i \mu_i = n$.

**Definition 3** (Minimum Size Coalescing Partition Problem)**.** Let $n$ and $k$ be integers, with $k < n$. We say that a partition $\mu$ of $n$ ***coalesces to*** a partition $\lambda$ of $n$ if it is possible to merge the terms of $\mu$ in a particular way so as to obtain $\lambda$: that is, there exists a decomposition of $\mu$ into disjoint subsets $\mu_{S_1}, \dots \mu_{S_{|\lambda|}}$ such that for each $i \in \{1, \dots, |\lambda|\}$, $\mu_{S_i} \vdash \lambda_i$. If for all $\lambda \vdash n$ such that $|\lambda| \leq k$, we have that $\mu$ can coalesce to $\lambda$, we will say that ***$\mu$ coalesces to all $k$-partitions of $n$***. The goal of the Minimum Size Coalescing Partition Problem (MSCP) is to find the $\mu$ of minimum size such that $\mu$ coalesces to all $k$-partitions of $n$.

As a simple example, if $n = 7$ and $k = 3$, then the solution to MSCP is the partition $\mu = \{3, 2, 1, 1\}$: there are a total of 8 partitions of the number 7 whose size at most 3, and $\mu$ coalesces to all of them:

1. $\{7\} = \{3 + 2 + 1 + 1\}$
2. $\{6, 1\} = \{3 + 2 + 1, 1\}$
3. $\{5, 2\} = \{3 + 2, 1 + 1\}$
4. $\{4, 3\} = \{3 + 1, 2 + 1\}$
5. $\{5, 1, 1\} = \{3 + 2, 1, 1\}$
6. $\{4, 2, 1\} = \{3 + 1, 2, 1\}$
7. $\{3, 3, 1\} = \{3, 2 + 1, 1\}$
8. $\{3, 2, 2\} = \{3, 2, 1 + 1\}$

In this case it is obvious that $\mu$ must be minimal because it has cardinality 4, and we certainly cannot hope to coalesce to all partitions of size at most 3 with any fewer than that.

The key result of this paper is a solution to MSCP for all $n$ and $k$. Let $n$ and $k$ be integers, with $k \leq n$. Let $\mu$ be constructed as follows. We will break off pieces of the total demand. For the first piece, $\mu_1$, take the ceiling of the total demand divided by $k$. For each subsequent piece take the ceiling of the remaining demand divided by $k$. We will prove this solution is optimal.

To prove this result we first show that the solution is feasible. We will make use of the following two key observations.

**Observation 4** (pigeonhole principle)**.** For every $\lambda \vdash n$ with $|\lambda| \leq k$ there exists some $i$ such that $\lambda_i \geq \lceil n/k \rceil$.

**Observation 5.** Let $\{\lambda_1, \ldots, \lambda_r\}$ be such that it is possible to complete the set $\{\lambda_1, \ldots, \lambda_r\}$ to a set that can coalesce to all $k$-partitions of $n$. If $\{\mu_1, \ldots, \mu_l\}$ coalesces to all $k$-partitions of $(n - \sum_{j=1}^{r} \lambda_j)$, then the partition $\{\lambda_1, \ldots, \lambda_r, \mu_1, \ldots, \mu_l\}$ is such a completion.

*Proof.* Suppose $\{\mu_1, \ldots, \mu_l\}$ coalesces to all $k$-partitions of $(n - \sum_{j=1}^{r} \lambda_j)$. Consider a partition $\gamma \vdash n$ with $|\gamma| \leq k$. Our assumption on the $\lambda_j$'s tells us that we can find a disjoint set decomposition $\cup_{i=1}^{|\gamma|} S_i = \{1, \ldots, r\}$ and $\beta_i, \ldots, \beta_{|\gamma|}$ such that

$$\gamma_i = \sum_{j \in S_i} \lambda_j + \beta_i \qquad \forall i \in \{1, \ldots, |\gamma|\}.$$

We must have $\beta \vdash (n - \sum_{j=1}^{r} \lambda_j)$. Thus $\mu$ coalesces to $\beta$, and we have a disjoint set decomposition $\cup_{i=1}^{k} T_i = \{1, \ldots, l\}$ for which

$$\gamma_i = \sum_{j \in S_i} \lambda_j + \sum_{j \in T_i} \mu_j \qquad \forall i \in \{1, \ldots, |\gamma|\},$$

or equivalently $\{\lambda_1, \ldots, \lambda_r, \mu_1, \ldots, \mu_l\}$ can coalesce to $\gamma$.

□

For any $n$, Observation 4 tells us that $\{\lceil n/k \rceil\}$ can always be completed to a set that coalesces to all $k$-partitions of $n$. In fact, $\lceil n/k \rceil$ is the largest value for which this is true, as adding any larger value would preclude coalescing to partitions with largest value $\lceil n/k \rceil$.

Our algorithm works greedily. It first makes $\mu_1$ the largest number such that it is always possible to complete $\{\mu_1\}$ to a coalescing set for all $k$-partitions of $n$. Observation 5 tells us that to complete our coalescing set we then need only find a coalescing set for all $k$-partitions of $n - \mu_1$. We make $\mu_2$ the largest number such that it would be possible to complete $\{\mu_2\}$ to a coalescing set for all $k$-partitions of $n - \mu_1$. To complete $\{\mu_2\}$ to our desired coalescing set we need only find a coalescing set for the all $k$-partitions of $n - (\mu_1 + \mu_2)$. We continue this until we consider $n = 1$ at which point the largest number such that we could complete the coalescing set itself completes the coalescing set, and we then have that the union of all of the $\{\mu_i\}$ must be able to coalesce to form all $k$-partitions of $n$.

We now prove that this solution is in fact optimal. We will rely on the following property of feasible MSCP solutions.

*Lemma* 1. Suppose $\lambda$ coalesces to all $k$-partitions of $n$. Then for any $\gamma \vdash n$ with $|\gamma| \leq k$, it is possible to form $\gamma$ greedily with $\lambda$ as follows: place $\lambda_1$ (the largest entry of $\lambda$) in the sum yielding the largest of the $\gamma_i$, that is $\gamma_1$. For each $j \in \{2, \ldots, |\lambda|\}$ iteratively place $\lambda_j$ in the sum yielding the largest $\gamma_i$ for which $\lambda_j$ can still fit in the sum.

*Proof.* Let $\gamma \vdash n$ with $|\gamma| \le k$. Suppose towards a contradiction that there is a point in our greedy formation of $\gamma$ at which we are trying to add a value $\lambda_{j^*}$, but it does not fit in any of the sums.

There is a disjoint set decomposition $\cup_{i=1}^{|\gamma|} S_i = \{1, \ldots, j^* - 1\}$ such that the state of our formation when we are trying to add $\lambda_{j^*}$ looks like

$$\gamma_i = \left( \sum_{j \in S_i} \lambda_j \right) + \alpha_i \qquad \forall i \in \{1, \ldots, |\gamma|\},$$

where $\alpha_i$ represents the amount of $\gamma_i$ we have yet to fill. We must have

$$\alpha_i < \lambda_{j^*} \qquad \forall i \in \{1, \ldots, |\gamma|\}.$$

The key to this argument is that, since $\lambda$ can be coalesced to form all $k$-partitions of $n$, it can be coalesced to form the following partition

$$\begin{aligned} &\alpha_{|\gamma|} \\ &\alpha_{|\gamma|-1} \\ &\vdots \\ &\alpha_2 \\ &n - \sum_{i=2}^{|\gamma|} \alpha_i. \end{aligned}$$

Furthermore, since each $\alpha_i$ is less than $\lambda_{j*}$, when we coalesce to form this partition we cannot use any of the values greater than or equal to $\lambda_{j^*}$ in the sums that yield the $\alpha_i$. Thus $\lambda_{j^*}$ and all of the values in $\lambda_j$ such that $j \in \cup_{i=1}^{|\gamma|} S_i$ must be used in the sum for $n - \sum_{i=2}^{|\gamma|} \alpha_i$. Thus

$$n - \sum_{i=2}^{|\gamma|} \alpha_i \ge \left( \sum_{i=1}^{|\gamma|} \sum_{j \in S_i} \lambda_j \right) + \lambda_{j^*}.$$

On the other hand we have

$$n = \left( \sum_{i=1}^{|\gamma|} \sum_{j \in S_i} \lambda_j \right) + \sum_{i=1}^{|\gamma|} \alpha_i.$$

Thus

$$n - \sum_{i=2}^{|\gamma|} \alpha_i = \left( \sum_{i=1}^{|\gamma|} \sum_{j \in S_i} \lambda_j \right) + \alpha_1.$$

Combining the above

$$\lambda_{j^*} \le \alpha_1,$$

a contradiction. Having arrived at a contradiction we conclude that at every step $j$ in our greedy generation we are able to add $\lambda_j$ to our sums. So we can introduce all the $\lambda_j$ without our sums ever exceeding the $\gamma_i$, which means that since introducing all of them introduces a total sum of $n$, we obtain all of $\gamma$.

$\square$

**Example 1.** For example, by our algorithm, (3,2,2,1,1) coalesces to all 3-partitions of 9. To form (4,3,2) we can use

$$(4,3,2) = (2+2, 3, 1+1),$$

but this lemma tells us we are also guaranteed to be able to use

$$(3+1, 2+1, 2).$$

The preceding allows us to show something akin to a converse of Observation 5:

*Lemma* 2. Let $\lambda = (\lambda_1 \geq \lambda_2 \geq \cdots \geq \lambda_l)$ be a partition that coalesces to all $k$-partitions of $\sum_{j=1}^{l} \lambda_j$. Then for all $m \in \{2, \dots, l\}$, we have

$$\{\lambda_m, \dots, \lambda_l\} \text{ can coalesce to all } k\text{-partitions of } \sum_{j=m}^{l} \lambda_j.$$

*Proof.* It clearly suffices to prove the claim for $m = 2$.

The proof is rather straightforward given Lemma 1. Suppose $\lambda$ coalesces to all $k$-partitions of $n$ with $|\lambda| = l$ and suppose $\nu$ is a partition of $(n - \lambda_1) = \sum_{j=2}^{l} \lambda_j$ with $|\nu| \leq k$. We construct a partition $\gamma$ of $n$ of the same size as $\nu$ by letting

$$\gamma_1 = \lambda_1 + \nu_1, \qquad \gamma_i = \nu_i \quad \forall i \in \{2, \dots, |\nu|\}.$$

Then $\gamma_1 = \lambda_1 + \nu_1$ is clearly the maximum element of $\gamma$. Thus by Lemma 1 it is possible to form $\gamma$ by coalescing $\lambda$ by first adding $\lambda_1$ to the sum for $\gamma_1$. Then the fact that we can fill out the rest of the $\gamma$ sums is precisely equivalent to we can obtain $\nu$ by coalescing $(\lambda_2, \dots, \lambda_l)$.

$\square$

To prove optimality, we rely on one more lemma.

*Lemma* 3. Let $\mu$ be the solution obtained by the algorithm. Let $\lambda$ be a partition that also coalesces to all $k$-partitions of $n$. Then

$$\sum_{i=1}^{m} \lambda_i \leq \sum_{i=1}^{m} \mu_i \qquad \forall m \in \{1, \dots, \min(|\mu|, |\lambda|)\}.$$

*Proof.* We induct on $m$. For $m = 1$ this is clear. If $\lambda_1 > \mu_1$ then $\lambda_1 > \lceil n/k \rceil$ which implies $\lambda_1$ cannot be used in forming any partitions $\gamma$ that have $\gamma_1 = \lceil n/k \rceil$. Thus $\lambda$ would not be feasible.

Now suppose the result holds for $m - 1$. Clearly

$$\sum_{i=1}^{m} \lambda_i = \left( \sum_{i=1}^{m-1} \lambda_i \right) + \lambda_m.$$

We know something about $\lambda_m$. By Lemma 2, $\lambda_m$ is the largest value in a partition that coalesces to all $k$-partitions of $n - \sum_{i=1}^{m-1} \lambda_i$. Thus $\lambda_m$ is less than or equal to the largest such a value could be, $\left\lceil (n - \sum_{i=1}^{m-1} \lambda_i)/k \right\rceil$. Thus

$$\begin{aligned}
\sum_{i=1}^{m} \lambda_i &= \left( \sum_{i=1}^{m-1} \lambda_i \right) + \lambda_m \\
&\leq \left( \sum_{i=1}^{m-1} \lambda_i \right) + \left\lceil \left( n - \sum_{i=1}^{m-1} \lambda_i \right) \Big/ k \right\rceil \\
&= \left\lceil \left( \sum_{i=1}^{m-1} \lambda_i \right) + \left( n - \sum_{i=1}^{m-1} \lambda_i \right) \Big/ k \right\rceil \\
&= \left\lceil \frac{n}{k} + \left( \frac{k-1}{k} \right) \sum_{i=1}^{m-1} \lambda_i \right\rceil \\
&\leq \left\lceil \frac{n}{k} + \left( \frac{k-1}{k} \right) \sum_{i=1}^{m-1} \mu_i \right\rceil \\
&= \left( \sum_{i=1}^{m-1} \mu_i \right) + \left\lceil \left( n - \sum_{i=1}^{m-1} \mu_i \right) \Big/ k \right\rceil \\
&= \left( \sum_{i=1}^{m-1} \mu_i \right) + \mu_m \\
&= \sum_{i=1}^{m} \mu_i.
\end{aligned}$$

□

We are now ready to prove the main theorem.

**Theorem 6.** *Let $n$ and $k$ be integers, with $k \leq n$. Let $\mu$ be constructed as follows:*

1. *Let $\mu_1 = \lceil n/k \rceil$.*
2. *For $i > 1$, let $\mu_i = \lceil (n - \sum_{j<i} \mu_j)/k \rceil$.*

*Then $\mu$ is a solution to MSCP: that is, $\mu$ is a minimum size partition that coalesces to all $k$-partitions of $n$.*

*Proof.* Let $\mu$ be the solution obtained from the algorithm. Suppose towards a contradiction that there exists a $\lambda^*$, with $|\lambda^*| < |\mu|$, that coalesces to all $k$-partitions of $n$.

By Lemma 3

$$\sum_{i=1}^{|\lambda^*|} \lambda_i^* \leq \sum_{i=1}^{|\lambda^*|} \mu_i < n.$$

a contradiction to $\lambda^*$ partitioning $n$. Having arrived at a contradiction we conclude the solution generated by the algorithm is optimal. □

The following corollary gives us a bound on the size of our coalescing partition.

**Corollary 7.** *Let $\mu$ be the minimum size partition that coalesces to all $k$-partitions of $n$. We have*

$$|\mu| \leq \lceil \log_{k/(k-1)}(n) \rceil + 1.$$

This upper bound becomes weak as $k$ increases relative to $n$, but we can quickly compute $|\mu|$ exactly for given $n$ and $k$ by simply running the algorithm. Table 1 gives the values of $|\mu|$ for $n \in \{1, \ldots, 20\}$, $k \in \{1, \ldots, 10\}$. We have also published these at the On-Line Encyclopedia of Integer Sequences [1, 2, 3, 4].

| $k$ \ $n$ | 1 | 2 | 3 | 4 | 5 | 6 | 7 | 8 | 9 | 10 | 11 | 12 | 13 | 14 | 15 | 16 | 17 | 18 | 19 | 20 |
|---|---|---|---|---|---|---|---|---|---|---|---|---|---|---|---|---|---|---|---|---|
| 1 | 1 | 1 | 1 | 1 | 1 | 1 | 1 | 1 | 1 | 1 | 1 | 1 | 1 | 1 | 1 | 1 | 1 | 1 | 1 | 1 |
| 2 | 1 | 2 | 2 | 3 | 3 | 3 | 3 | 4 | 4 | 4 | 4 | 4 | 4 | 4 | 4 | 5 | 5 | 5 | 5 | 5 |
| 3 | 1 | 2 | 3 | 3 | 4 | 4 | 4 | 5 | 5 | 5 | 5 | 6 | 6 | 6 | 6 | 6 | 6 | 7 | 7 | 7 |
| 4 | 1 | 2 | 3 | 4 | 4 | 5 | 5 | 6 | 6 | 6 | 7 | 7 | 7 | 7 | 8 | 8 | 8 | 8 | 8 | 9 |
| 5 | 1 | 2 | 3 | 4 | 5 | 5 | 6 | 6 | 7 | 7 | 7 | 8 | 8 | 8 | 9 | 9 | 9 | 9 | 10 | 10 |
| 6 | 1 | 2 | 3 | 4 | 5 | 6 | 6 | 7 | 7 | 8 | 8 | 9 | 9 | 9 | 10 | 10 | 10 | 11 | 11 | 11 |
| 7 | 1 | 2 | 3 | 4 | 5 | 6 | 7 | 7 | 8 | 8 | 9 | 9 | 10 | 10 | 10 | 11 | 11 | 11 | 12 | 12 |
| 8 | 1 | 2 | 3 | 4 | 5 | 6 | 7 | 8 | 8 | 9 | 9 | 10 | 10 | 11 | 11 | 12 | 12 | 12 | 13 | 13 |
| 9 | 1 | 2 | 3 | 4 | 5 | 6 | 7 | 8 | 9 | 9 | 10 | 10 | 11 | 11 | 12 | 12 | 13 | 13 | 13 | 14 |
| 10 | 1 | 2 | 3 | 4 | 5 | 6 | 7 | 8 | 9 | 10 | 10 | 11 | 11 | 12 | 12 | 13 | 13 | 14 | 14 | 15 |

Table 1: The size of the coalescing partition $\mu$ obtained from our solution for a range of $n$ and $k$. Since we show our solution is optimal this is the minimum size needed to be able to obtain all $\lambda \vdash n$ with $|\lambda| \leq k$ by coalescing parts of $\mu$.

*Proof of Corollary 7.* Let $L(n, k)$ be the size, i.e. number of nonzero terms, of the $\mu$ constructed in this way for inputs $n$ and $k$. We do not have a closed form expression for $L(n, k)$, but it is clear that we have the following recurrence relation

$$L(n, k) = 1 + L(n - \lceil n/k \rceil, k), \quad L(0, k) = 0,$$

equivalently

$$L(n, k) = 1 + L\left(\left\lfloor \left(\frac{k-1}{k}\right) n \right\rfloor, k\right), \quad L(1, k) = 1.$$

Let $L'(n, k)$ satisfy the recurrence relation

$$L'(n, k) = 1 + L\left(\left(\frac{k-1}{k}\right) n, k\right), \quad L'(1, k) = 1.$$

Then clearly for any fixed $n, k$, we have $L'(n, k) \geq L(n, k)$. In addition $L'(n, k)$ is increasing in $n$.

Thus $L(n, k)$ must have value less than or equal to that of

$$L'\left(\left(\frac{k}{k-1}\right)^{\lceil \log_{k/(k-1)}(n) \rceil}, k\right) = \lceil \log_{k/(k-1)}(n) \rceil + 1.$$

We thus have $|\mu| \leq \lceil \log_{k/(k-1)}(n) \rceil + 1$. □

The preceding analysis describes a splitting rule that addresses the trade-off of size and quality in a theoretically sound way. Our splitting rule simply takes each $d_i$ and splits it into the pieces given by minimum size partition that coalesces to all $k$-partitions of $d_i$, constructed precisely as in Theorem 6. Now, all possible splits of each customer demand over different trucks can be manifested in the CVRP solution by trucks visiting some combination of our pieces, and therefore, all possible splits can be recovered in a SDVRP solution. We throw out no feasible solutions, and thus if we optimally solve the CVRP, we have optimally solved the SDVRP. Furthermore, Corollary 7 gives us a closed form upper bound on the size of our reduction,

$$\text{Number of CVRP nodes} \leq \sum_{i=1}^{n} (\lceil \log_{k/(k-1)}(d_i) \rceil + 1),$$

though from observation (see Table 1), we expect the size to in fact be much smaller.

**Observation 8.** In practice, it is often sensible to assume that the actual amount of demand splitting between trucks is not too large; in other words, there is likely some $\bar{k} < k$ for which we can expect optimal solutions to use at most $\bar{k}$ trucks to visit any given customer (in plain and simple English: we might have $k = 100$ vehicles at our disposal, but deem it unlikely that any demand node is serviced by more than (say) $\bar{k} = 3$ vehicles). Therefore, instead of requiring our splitting rule to coalesce to all $k$-partitions, we can instead coalesce to all $\bar{k}$-partitions. The result is that our splitting rule still allows us to find any feasible solutions that would conform to our simplifying assumption, and by making this modification we have reduced the size of the CVRP problem that we need to solve.

**Observation 9.** In order to apply an a priori splitting rule it must be the case that our demands are integers and such an integral amount of our commodity can be divided into discrete size-1 parts. It is possible to discretize the division of a continuous commodity by taking a size-1 part to be a small amount of that commodity and only allowing splits to take on multiples of this amount. A granularity of the discretization must then be chosen. Corollary 7 tells us that choosing a sufficiently small size-1 amount of our commodity need not present a problem in terms of blowing up the size of the resulting a priori split problem. Halving the size of the amount that constitutes a unit doubles each demand, but Corollary 7 tells us, letting $a = k/(k-1)$, the new size of our minimum size coalescing partition is at most

$$1 + \lceil \log_a(2d) \rceil \leq 1 + \lceil \log_a(d) \rceil + \lceil \log_a(2) \rceil,$$

i.e. our bound on the partition size for demand $d$ plus $\lceil \log_a(2) \rceil$. In multiplying our granularity we have at most *added* to the problem size a constant independent of the original size of the demands.

# 4 Computational results

In the previous section, it is elaborated that our *a priori* splitting rule enables the reduction of the split-delivery vehicle routing problem (SDVRP) into the capacitated vehicle routing problem (CVRP) through the simple division of customer nodes into several nodes with reduced demands. This process increases the problem size yet maintains the lossless property, allowing the utilization of existing CVRP solvers for the SDVRP. This section undertakes computational experiments on the benchmarks of the SDVRP and its time window variant (SDVRPTW), leveraging the enumeration schemes that optimally solve the minimum size coalescing partition problem.

## 4.1 Instances

The LKH-3 solver [23] has demonstrated effectiveness in tackling constrained vehicle routing problems including the CVRP and CVRPTW, making it the solver of choice for the CVRP subproblems generated by the lossless *a priori* splitting rule. The results obtained are competitive with the current state-of-the-art for both the SDVRP and SDVRPTW, achieved without significant adjustments to the metaheuristics. Notably, when comparing the computational outcomes in this section with those derived from other *a priori* splitting rules, such as the strategy developed by [17], the method in our work exhibits a marked advantage.

The performance evaluation of our proposed method spans two sets of SDVRP benchmarks and one set of SDVRPTW benchmarks. The first set, presented by [13], includes 25 instances. This set features the exact Christofides and Eilon instances from TSPLIB [29] and additional instances with the same coordinates as eil51, eil76, eil101, and a fixed vehicle capacity of $Q = 160$. Customer demands across these instances are randomly generated according to various ranges, proportionally scaled to the vehicle's capacity: D1: $[\lceil 0.01Q \rceil, \lfloor 0.1Q \rfloor]$; D2: $[\lceil 0.1Q \rceil, \lfloor 0.3Q \rfloor]$; D3: $[\lceil 0.1Q \rceil, \lfloor 0.5Q \rfloor]$; D4: $[\lceil 0.1Q \rceil, \lfloor 0.9Q \rfloor]$; D5: $[\lceil 0.3Q \rceil, \lfloor 0.7Q \rfloor]$; D6: $[\lceil 0.7Q \rceil, \lfloor 0.9Q \rfloor]$.

The second set of benchmarks, devised by [11], comprises 42 instances, each applying varied demand generation strategies across six different customer maps, distinguished by the diversity in customer count and location. The first and second sets of SDVRP benchmark instances have been extensively examined in a wide range of researches [5, 6, 13, 17, 22, 27, 32], with [22, 32] achieving most of the best known solutions.

The third set introduces SDVRPTW benchmark instances, initially proposed by [33] and categorized into groups C, RC, and R. The computational experiments detailed in this section focus on the C and RC groups, featuring 58 instances notable for lacking feasible non-split solutions. The best known SDVRPTW solutions for these instances have been recognized and documented by [7, 20, 25].

Preliminary computational experiments were conducted on the classical benchmark Set 1, as introduced by [13]. These experiments aim to evaluate the impact of constraining the sizes of CVRP subproblems generated by the *a priori* splitting rule on both the computational time and the quality of solutions for the SDVRP. Furthermore, the effectiveness of the splitting rule in tackling more complex split-delivery routing challenges was further confirmed through a comparative analysis with the leading solutions for the SDVRP problems in Set 2 and the SDVRPTW problems in Set 3.

The LKH-3 solver was executed 10 times for each problem instance, employing non-default hyperparameter settings labeled "SPECIAL" and modifying the maximum number of trials in each run through the hyperparameter "MAX_TRIALS". Details can be found in the description of parameters for LKH-3 [23]. By standard configuration, LKH-3 addresses the split-delivery vehicle routing problem with a limited fleet (SDVRP-LF) rather than the variant with an unlimited fleet (SDVRP-UF), setting the fleet size to the smallest feasible number of vehicles, denoted as $k_{min} = \lceil \sum_{i=1}^{n} d_i/Q \rceil$. Consequently, for the first and second instance sets, the computational outcomes are benchmarked against the best known SDVRP-LF solutions. For the third instance set, concerning split-delivery routing problems with time windows, the derived SDVRPTW-LF solutions are evaluated in comparison to the best known SDVRPTW solutions documented by [7, 20, 25].

All experiments were conducted using Intel Xeon E5-2640 v3, 2.60 GHz CPUs at the USC Center for Advanced Research Computing (CARC), equipped with 59 GB of memory per node. The computational time was limited to three hours for instances from [13], and extended to six hours for instances from [11, 33].

## 4.2 Limiting the non-split problem size

In the tables presented from now on, several key terms are defined for clarity. *Instance* refers to the name of the problem instance. $k$ represents the minimum feasible number of vehicles, denoted before as $k_{min}$. *BKS* stands for the best known solution for the SDVRP (or SDVRPTW) benchmark instance. *Best sol.* indicates the best solution identified by the *a priori* splitting rule for the SDVRP (or SDVRPTW) across all executions. *Gap (%)* is the percentage difference between the best solution and the BKS, calculated as $100 \times (\textit{Best sol.} - \textit{BKS})/\textit{BKS}$. *Time (s)* denotes the time in seconds to achieve the best solution. Solutions that equal or surpass the BKS will be highlighted in bold.

Building on earlier observations, it is postulated that the extent of demand splitting across trucks is likely not excessively large in practical scenarios. It is generally expected that some $\bar{k}$, significantly lower than $k$, exists, ensuring that no more than $\bar{k}$ trucks are required to service any specific customer. This premise facilitates a significant reduction in the sizes of the generated non-split CVRP instances requiring solutions.

Meanwhile, the hyperparameter $q$ dictates that only customer demands greater than $qQ$ should be considered for division using the minimum size coalescing partition strategy. This approach stems from the understanding that segmenting small customer demands into smaller units does not typically enhance solution quality but tends to adversely affect computational time. When $q = 1$ (or more precisely when $q \geq d_{max}/Q$), the scenario simplifies to that of a non-split capacitated vehicle routing problem instance. Conversely, when $q = 0$ (or more precisely when $q < d_{min}/Q$), the scenario of a fully relaxed split-delivery vehicle routing problem needs to be addressed.

In the forthcoming Tables 2 and 3, the columns for *Best* $\bar{k}$ and *Best* $q$ are included to showcase the effects of $\bar{k}$ and $q$ on both the solution quality and computation time of the generated non-split CVRP instances. Specifically, $\bar{k}$ and $q$ are defined within the domains $\bar{k} \in \{2, \ldots, k\}$ and $q \in [0, 1]$, respectively. A lesser value of $\bar{k}$ yields a reduced size for the generated non-split problem, and similarly, an increased $q$ value also diminishes the generated non-split problem size. Overall, this measure demonstrates how $\bar{k}$ and $q$ can be effectively utilized to reduce the non-split problem size, in comparison to completing the minimum size coalescing partitioning procedure, while maintaining or enhancing the solutions for SDVRP benchmark instances.

Lastly, the *Full size* column in Tables 2 and 3 refers to the comprehensive magnitude of the generated non-split problem that guarantees a lossless solution to each split-delivery routing problem instance. Setting $\bar{k} = k$ and $q = 0$, it represents the aggregate cardinality of the minimum size partitions $\mu_i$ that coalesce to all $k$-partitions of each customer demand $d_i$. *Prob size* signifies the size of the non-split problem created with the specified $\bar{k}$ and $q$, which facilitate the best split-delivery solution identified by the LKH-3 solver. *Ratio (%)* denotes the percentage ratio of the problem size that culminates in the best outcome, relative to the comprehensive non-split problem size, computed as 100 × *Prob size*/*Full size*.

| Instance | $k$ | BKS | Best sol. | Gap (%) | Time[a] (s) | Prob size | Full size | Ratio (%) | Best $\bar{k}$ | Best $q$ |
|---|---|---|---|---|---|---|---|---|---|---|
| eil22 | 4 | 375[b] | **375** | **0.000** | 0.01 | 33 | 455 | 7.25 | 2 | 0.40 |
| eil23 | 3 | 569[c] | **569** | **0.000** | 0.01 | 42 | 287 | 14.63 | 3 | 0.90 |
| eil30 | 3 | 510[b] | **510** | **0.000** | 0.02 | 64 | 381 | 16.80 | 3 | 0.30 |
| eil33 | 4 | 835[c] | **835** | **0.000** | 0.07 | 44 | 654 | 6.73 | 2 | 0.45 |
| eil51 | 5 | 521[b] | **521** | **0.000** | 0.43 | 59 | 417 | 14.15 | 3 | 0.25 |
| eilA76 | 10 | 818[c] | **818** | **0.000** | 308.30 | 465 | 989 | 47.02 | 4 | 0.10 |
| eilB76 | 14 | 1002[c] | 1003 | 0.100 | 30.83 | 187 | 1137 | 16.45 | 2 | 0.20 |
| eilC76 | 8 | 733[c] | **733** | **0.000** | 86.44 | 322 | 887 | 36.30 | 2 | 0.05 |
| eilD76 | 7 | 681[c] | 682 | 0.147 | 0.11 | 90 | 814 | 11.06 | 3 | 0.15 |
| eilA101 | 8 | 815[c] | **815** | **0.000** | 0.59 | 137 | 1005 | 13.63 | 3 | 0.15 |
| eilB101 | 14 | 1061[c] | 1062 | 0.094 | 269.18 | 427 | 1246 | 34.27 | 3 | 0.10 |
| S51D1 | 3 | 458[c] | **458** | **0.000** | 0.23 | 167 | 220 | 75.91 | 2 | 0.00 |
| S51D2 | 9 | 703[c] | 705 | 0.284 | 34.99 | 381 | 788 | 48.35 | 3 | 0.10 |
| S51D3 | 15 | 942[e] | 945 | 0.318 | 10.31 | 265 | 1288 | 20.57 | 3 | 0.25 |
| S51D4 | 27 | 1551[e] | 1560 | 0.580 | 10584.38 | 842 | 2381 | 35.36 | 6 | 0.20 |
| S51D5 | 23 | 1328[c] | 1334 | 0.452 | 528.29 | 745 | 2092 | 35.61 | 8 | 0.35 |
| S51D6 | 41 | 2153[d] | 2183 | 1.393 | 2635.05 | 1035 | 3804 | 27.21 | 6 | 0.70 |
| S76D1 | 4 | 592[c] | **592** | **0.000** | 17.67 | 292 | 400 | 73.00 | 4 | 0.05 |
| S76D2 | 15 | 1081[c] | 1086 | 0.463 | 848.87 | 413 | 1646 | 25.09 | 2 | 0.05 |
| S76D3 | 23 | 1419[c] | 1427 | 0.564 | 1595.84 | 611 | 2446 | 24.98 | 4 | 0.20 |
| S76D4 | 37 | 2071[c] | 2085 | 0.676 | 1726.02 | 624 | 3852 | 16.20 | 3 | 0.20 |
| S101D1 | 5 | 716[c] | **716** | **0.000** | 40.42 | 326 | 567 | 57.50 | 2 | 0.00 |
| S101D2 | 20 | 1360[e] | 1375 | 1.103 | 473.32 | 542 | 2468 | 21.96 | 2 | 0.10 |
| S101D3 | 31 | 1858[e] | 1874 | 0.861 | 2001.65 | 1091 | 3772 | 28.92 | 7 | 0.25 |
| S101D5 | 48 | 2767[e] | 2819 | 1.879 | 5832.25 | 563 | 6168 | 9.13 | 3 | 0.45 |
| Average | | | | 0.357 | 1081.01 | | | 28.72 | 3.4 | 0.24 |

[a] Intel Xeon E5-2640 v3, 2.6 GHz
[b] Belenguer et al. (2000) [13]
[c] Silva et al. (2015) [32]
[d] Ozbaygin et al. (2018) [27]
[e] He and Hao (2023) [22]

Table 2: LKH-3 results for SDVRP with rounded costs and limited fleet on the instances of Belenguer et al. [13]

Upon reviewing all 25 instances from Set 1, Figures 1 and 2 depict how different choices of $\bar{k}$ and $q$ influence the size of the non-split problem and the gap of the LKH-3 solution from the best known solution, respectively. The heat map in Figure 2 reveals that the solution gap tends to be higher in the bottom right corner (where $\bar{k}$ is large and $q$ is small) and at the top (where $q$ is large). This outcome is logical, as an excessively large problem size may lead to diminished solution quality due to the constraints of limited computation time. In contrast, a higher $q$ limits the permissible splitting, producing a capacitated vehicle routing problem that is more akin to the problem instance which prohibits splitting. Thus, the ideal selections for $q$ appear to be neither excessively small (approaching 0) nor excessively large (approaching $d_{max}/Q$), but rather an intermediate value.

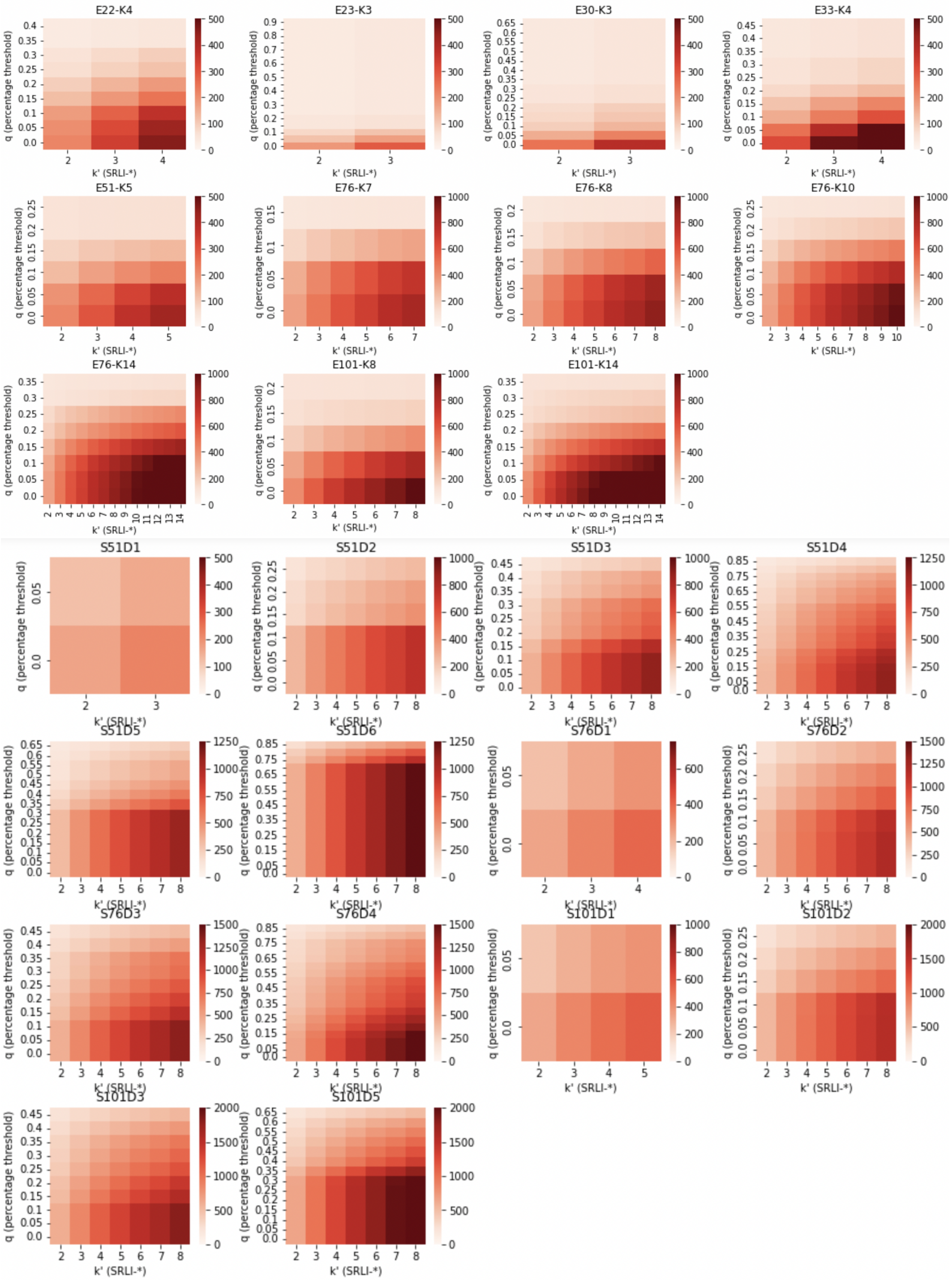

Figure 1: The *a priori* splitting rule, when applied to Instance Set 1 from [13], results in varying sizes of non-split capacitated vehicle routing problem instances for different $(\bar{k}, q)$ combinations. The generated problem size increases with larger $\bar{k}$ or smaller $q$. It is important to note that when $q \geq d_{max}/Q$, the resulting problems effectively disallow splitting, hence such instances are excluded from Figure 1.

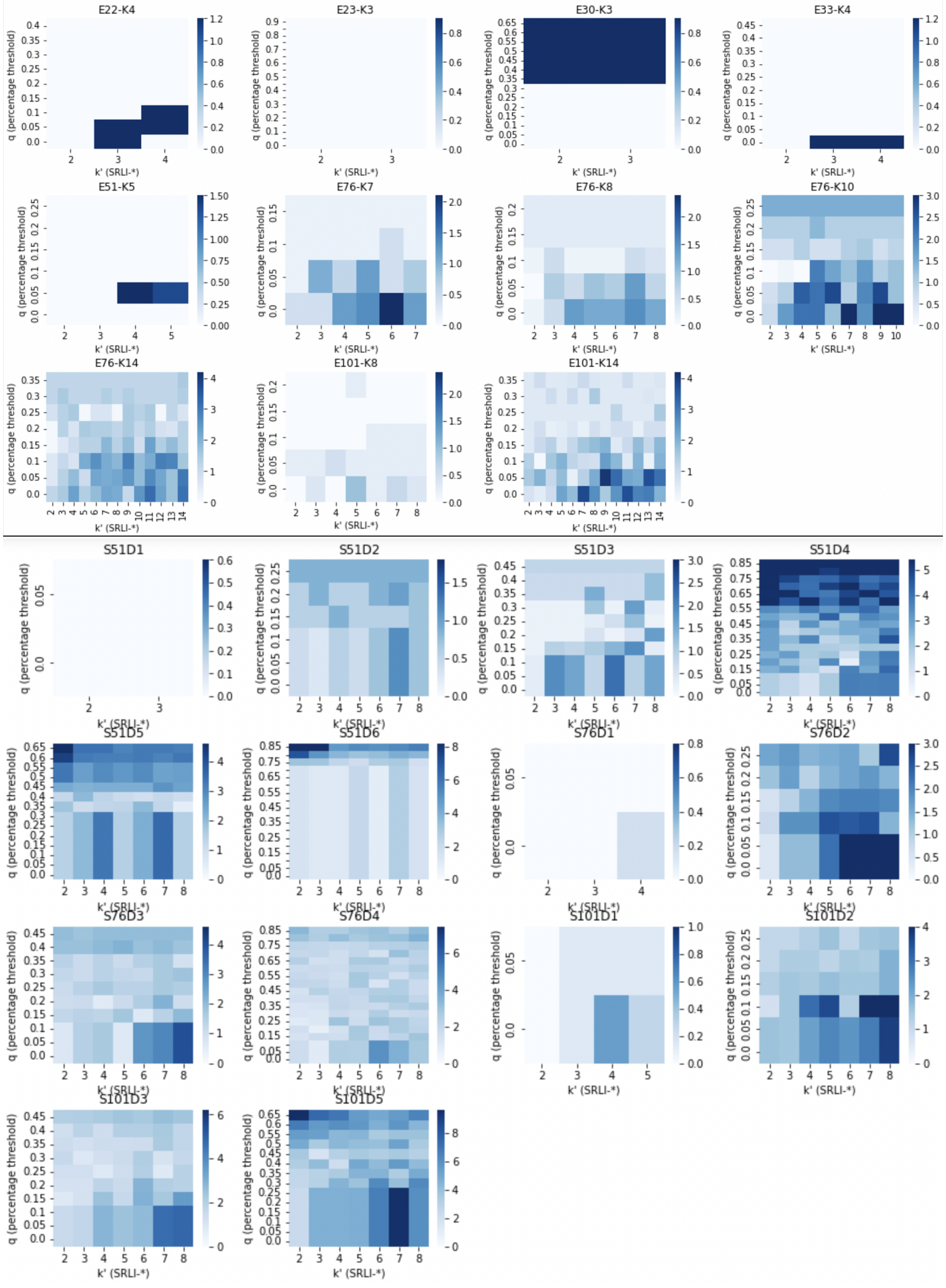

Figure 2: The gap between the solution obtained using the *a priori* splitting rule and the best known solution, for varying selections of $(\bar{k}, q)$, analyzed across all instances of Set 1 [13].

Table 2 focuses on the 25 Set 1 instances introduced by [13], showcasing the results of applying the lossless *a priori* splitting rule, in conjunction with the LKH-3 solver. This analysis considers all practical $q$ values, initiating at 0 and increasing in increments of 0.05, up to 1. For instances where different $q$ values result in the same non-split problem, the splitting rule utilizes the highest $q$ value. The range of $\bar{k}$ values analyzed is $\{2, 3, \ldots, 8\}$, based on the rationale that allocating more than this number of vehicles to service the same customer is unlikely in practical scenarios.

Findings on the selection of $\bar{k}$ and $q$ are also presented in Table 2. The evaluation of different ($\bar{k}$, $q$) combinations prioritizes solution quality, followed by computational runtime. For each instance, Table 2 details the ($\bar{k}$, $q$) pair leading to the best solution as determined by the LKH-3 solver, along with the full and reduced non-split problem sizes, as well as the reduction ratio. It is observed that in most scenarios, $\bar{k} = 2$ and $\bar{k} = 3$ yield the ideal outcomes, surpassing those obtained with larger values of $\bar{k}$, which correspond to increased non-split problem sizes. This observation aligns with the rationale that a small number of vehicles should suffice for serving each customer. Moreover, in most instances, a nonzero value of $q$ is chosen to limit the number of demand nodes that require splitting, thereby diminishing the overall problem size. The average selections for $\bar{k}$ and $q$ are found to be 3.4 and 0.24, respectively. The average reduction in problem size across the 25 instances is 28.72%, signifying that the problem size can be reduced to 28.72% relative to the solution to the minimum size coalescing partition (MSCP), without adversely affecting the solution quality of the SDVRP.

The solutions reported in Table 2 demonstrate a match with 11 of the 25 best known solutions for the SDVRP, achieved through the application of the splitting rule alone, without any advanced adjustments to the LKH-3 metaheuristics. The average gap between these solutions and the BKS is a modest 0.357%. Note that the focus of this paper is not to modify the metaheuristics but to demonstrate the efficacy of the *a priori* splitting rule. The LKH-3 solver, not being able to divide non-split instances further into separate subtasks for multi-core processor utilization, is optimized for single-thread execution. Consequently, all computational experiments were conducted using a single thread. Despite these constraints, computational experiments in this section have achieved solution quality comparable to that reported in [22, 32] within a feasible computation time.

Comprehensive computational experiments were also conducted on the 42 instances of Set 2, as introduced by [11], with findings summarized in Table 3. It is noted that the minimum coalescing partition sizes for this set are considerably larger on average compared to those in Set 1. Considering the excessive sizes of non-split problems, evaluations were limited to configurations with $\bar{k} \in \{2, 3\}$ and $q \in \{0, 0.05, 0.2, 0.3, 0.5, 0.8\}$. The average gap from the BKS identified by [22, 32] is 1.122%. Note that the contributions to metaheuristic development for split-delivery routing problems by [22, 32] are widely recognized, whereas this paper underscores the remarkable efficacy of enumeration schemes in addressing the SDVRP, facilitated by *a priori* splitting rules. Achieving solution quality on par with the state-of-the-art is deemed satisfactory.

| Instance | $k$ | BKS | Best sol. | Gap (%) | Time[a] (s) | Prob size | Full size | (%) | $\bar{k}$ | $q$ |
|---|---|---|---|---|---|---|---|---|---|---|
| p01-50 | 5 | 524.61[b] | **524.61** | **0.000** | 0.02 | 56 | 417 | 13.43 | 2 | 0.20 |
| p02-75 | 10 | 823.89[b] | 831.79 | 0.959 | 1413.20 | 478 | 989 | 48.33 | 3 | 0.00 |
| p03-100 | 8 | 826.14[b] | **826.14** | **0.000** | 0.51 | 109 | 1005 | 10.85 | 3 | 0.20 |
| p04-150 | 12 | 1023.23[c] | 1029.56 | 0.619 | 23.20 | 167 | 1817 | 9.19 | 3 | 0.20 |
| p05-199 | 16 | 1285.79[b] | 1295.96 | 0.791 | 17493.23 | 216 | 2814 | 7.68 | 3 | 0.20 |
| p11-120 | 7 | 1037.88[b] | 1042.12 | 0.409 | 0.42 | 121 | 1029 | 11.76 | 2 | 0.20 |
| p01-50D1 | 3 | 459.50[b] | **459.50** | **0.000** | 0.21 | 126 | 228 | 55.26 | 2 | 0.05 |
| p02-75D1 | 5 | 617.85[b] | 621.60 | 0.607 | 1.28 | 193 | 425 | 45.41 | 2 | 0.05 |
| p03-100D1 | 6 | 760.00[b] | 760.19 | 1.006 | 3161.28 | 313 | 803 | 38.98 | 2 | 0.05 |
| p04-150D1 | 9 | 921.20[c] | 925.99 | 0.520 | 5315.51 | 466 | 1440 | 32.36 | 2 | 0.05 |
| p05-199D1 | 12 | 1073.57[c] | 1086.14 | 1.171 | 10042.39 | 769 | 2021 | 38.05 | 3 | 0.05 |
| p11-120D1 | 8 | 1042.80[c] | 1042.89 | 0.009 | 1819.58 | 475 | 1103 | 43.06 | 2 | 0.00 |
| p01-50D2 | 11 | 756.71[b] | 768.79 | 1.596 | 216.22 | 278 | 952 | 29.20 | 2 | 0.00 |
| p02-75D2 | 16 | 1109.24[c] | 1117.34 | 0.730 | 370.51 | 265 | 1628 | 16.28 | 2 | 0.20 |
| p03-100D2 | 22 | 1458.46[b] | 1466.20 | 0.531 | 2593.03 | 565 | 3116 | 18.13 | 3 | 0.20 |
| p04-150D2 | 32 | 2016.93[c] | 2051.91 | 1.734 | 14050.17 | 596 | 5442 | 10.95 | 2 | 0.20 |
| p05-199D2 | 41 | 2478.37[c] | 2508.80 | 1.228 | 20282.88 | 1148 | 7656 | 14.99 | 2 | 0.00 |
| p11-120D2 | 26 | 2898.25[c] | 2906.88 | 0.298 | 16325.68 | 681 | 4029 | 16.90 | 3 | 0.20 |
| p01-50D3 | 16 | 1005.75[b] | 1014.91 | 0.911 | 5323.18 | 304 | 1410 | 21.56 | 2 | 0.00 |
| p02-75D3 | 24 | 1502.05[b] | 1517.66 | 1.039 | 17693.39 | 637 | 2437 | 26.14 | 3 | 0.00 |
| p03-100D3 | 33 | 1996.76[c] | 2021.21 | 1.224 | 4413.62 | 638 | 4663 | 13.68 | 3 | 0.30 |
| p04-150D3 | 49 | 2849.59[c] | 2901.45 | 1.820 | 16784.15 | 688 | 8234 | 8.36 | 2 | 0.30 |
| p05-199D3 | 63 | 3469.90[c] | 3525.17 | 1.593 | 20120.86 | 855 | 11471 | 7.45 | 2 | 0.30 |
| p11-120D3 | 40 | 4215.98[c] | 4236.70 | 0.491 | 21332.67 | 770 | 6103 | 12.62 | 3 | 0.30 |
| p01-50D4 | 26 | 1487.18[c] | 1496.13 | 0.602 | 17425.97 | 332 | 2303 | 14.42 | 2 | 0.00 |
| p02-75D4 | 40 | 2301.61[b] | 2319.73 | 0.787 | 5149.12 | 488 | 4030 | 12.11 | 2 | 0.00 |
| p03-100D4 | 56 | 3085.69[c] | 3120.80 | 1.138 | 20732.48 | 721 | 7815 | 9.23 | 2 | 0.00 |
| p04-150D4 | 83 | 4533.82[c] | 4660.13 | 2.786 | 21381.17 | 1109 | 13814 | 8.03 | 3 | 0.50 |
| p05-199D4 | 105 | 5521.61[c] | 5654.87 | 2.413 | 21437.06 | 1368 | 19020 | 7.19 | 3 | 0.50 |
| p11-120D4 | 67 | 6849.73[c] | 6974.97 | 1.828 | 15613.14 | 884 | 10150 | 8.71 | 3 | 0.50 |
| p01-50D5 | 26 | 1481.71[b] | 1487.81 | 0.412 | 992.34 | 506 | 2371 | 21.34 | 3 | 0.00 |
| p02-75D5 | 39 | 2219.11[c] | 2260.89 | 1.883 | 21538.35 | 732 | 4083 | 17.93 | 3 | 0.00 |
| p03-100D5 | 53 | 2986.27[c] | 3025.42 | 1.311 | 16483.32 | 473 | 7700 | 6.14 | 2 | 0.50 |
| p04-150D5 | 79 | 4332.75[c] | 4400.41 | 1.562 | 18464.32 | 1051 | 13637 | 7.71 | 3 | 0.50 |
| p05-199D5 | 102 | 5398.15[c] | 5521.30 | 2.281 | 20779.62 | 890 | 19384 | 4.59 | 2 | 0.50 |
| p11-120D5 | 64 | 6639.95[c] | 6712.05 | 1.086 | 10921.86 | 831 | 10095 | 8.23 | 3 | 0.50 |
| p01-50D6 | 41 | 2155.80[c] | 2172.10 | 0.756 | 20473.01 | 378 | 3793 | 9.97 | 2 | 0.00 |
| p02-75D6 | 61 | 3217.51[c] | 3257.50 | 1.243 | 18018.49 | 526 | 6493 | 8.10 | 2 | 0.00 |
| p03-100D6 | 82 | 4378.33[c] | 4453.35 | 1.713 | 19439.10 | 1201 | 12116 | 9.91 | 3 | 0.00 |
| p04-150D6 | 122 | 6378.28[c] | 6538.92 | 2.519 | 20736.82 | 1201 | 21278 | 5.64 | 2 | 0.00 |
| p05-199D6 | 161 | 8181.44[c] | 8389.22 | 2.540 | 20880.95 | 1593 | 31536 | 5.05 | 2 | 0.00 |
| p11-120D6 | 98 | 10192.00[c] | 10291.36 | 0.975 | 21157.62 | 961 | 15588 | 6.16 | 2 | 0.00 |
| Average | | | | 1.122 | 11676.24 | | | 17.17 | 2.43 | 0.16 |

[a] Intel Xeon E5-2640 v3, 2.6 GHz
[b] Silva et al. [32]
[c] He and Hao [22]

Table 3: LKH-3 results for SDVRP with limited fleet on the instances of Archetti et al. [11]

The result of experiments further indicates a slight preference for $\bar{k} = 2$ over $\bar{k} = 3$. The average reduction in problem size for the 42 instances in Set 2 is 17.17%, demonstrating that by permitting two or three trucks to serve each customer, the problem size can be effectively reduced to 17.17% of the minimum coalescing partition size.

### 4.3 Adding time windows

This section of the paper delves into the split-delivery vehicle routing problem with time windows (SDVRPTW), as delineated in Definition 2. The investigation employs the *a priori* splitting rule on a collection of classic instances introduced by [33], specifically focusing on 58 instances from groups C and RC, with sizes $n = 26, 51$ and vehicle capacity $Q = 30$. Instances with the same prefix have uniform node coordinates and demands, but exhibit substantial variations in time windows at the individual instance level. Notably, all instances feature at least one node with a demand exceeding the vehicle's capacity, precluding any feasible non-split solutions and necessitating splitting in the split-delivery context. The best known solutions for these instances were achieved by [7, 20, 25].

In Table 4, the column titled *No. split* reflects the number of splits occurring within the chosen tour generated by LKH-3. For the 58 instances of Set 3, the average gap between the outcomes produced by the *a priori* splitting rule in conjunction with the LKH-3 algorithm and the BKS is a negligible 0.084%, with the methodology of this paper successfully matching 32 of the BKS within Group RC. For the 26 instances categorized under Group C, the average gap slightly increases to 0.112%.

For each instance in Set 3, binary decisions were made regarding $\bar{k} \in \{2, 3\}$ and $q \in \{0, 0.5\}$. The mean value of $\bar{k}$ calculated is 2.14, indicating a predominant preference for $\bar{k} = 2$ over $\bar{k} = 3$. The average value of $q$ stands at 0.44, suggesting a general inclination towards $q = 0.5$ rather than $q = 0$. The mean number of splits in the solutions that were realized for split-delivery is 4.54, confirming that splits do occur. The average reduction in problem size is 21.72%, showcasing the effective decrease relative to the minimum coalescing partition size, while still maintaining solution quality.

## 5 Conclusions

We have introduced a heuristic for split-delivery routing problems. Building on the work of [17], we developed a demand splitting rule that yields the minimum-size reduction of a split-delivery instance of a problem to an equivalent unsplittable instance. Inherent to our method is the ability to incorporate an assumed upper bound on the number of vehicles servicing a given customer (that is, $k$ or $\bar{k}$), thereby further reducing the size of the resulting non-split problem while preserving solution quality. Our computational experiments for the Split-Delivery Vehicle Routing Problem (SDVRP) show that our splitting rule comes close to the performance of state of the art algorithms. Its real value is demonstrated in its immediate applicability to other extensions to the vehicle routing problem with split delivery. Our computational experiments for the Split-Delivery Vehicle Routing Problem with Time Windows show comparable performance to the state-of-the-art.

| Instance | $k$ | BKS | Best sol. | Gap (%) | Time[a] (s) | Prob size | (%) | No. split | $\bar{k}$ | $q$ |
|---|---|---|---|---|---|---|---|---|---|---|
| C101.25.30 | 16 | 821.1[b] | 821.3 | 0.024 | 1.77 | 109 | 28.02 | 3 | 3 | 0.50 |
| C102.25.30 | 16 | 821.1[c] | 821.3 | 0.024 | 0.40 | 81 | 20.82 | 3 | 2 | 0.50 |
| C103.25.30 | 16 | 821.1[c] | 821.3 | 0.024 | 0.31 | 81 | 20.82 | 3 | 2 | 0.50 |
| C104.25.30 | 16 | 821.1[c] | 821.3 | 0.024 | 0.18 | 81 | 20.82 | 3 | 2 | 0.50 |
| C105.25.30 | 16 | 821.1[b] | 821.3 | 0.024 | 82.69 | 81 | 20.82 | 3 | 2 | 0.50 |
| C106.25.30 | 16 | 821.1[b] | 821.3 | 0.024 | 3.48 | 117 | 30.08 | 3 | 2 | 0.00 |
| C107.25.30 | 16 | 821.1[c] | 821.3 | 0.024 | 0.70 | 81 | 20.82 | 3 | 2 | 0.50 |
| C108.25.30 | 16 | 821.1[c] | 821.3 | 0.024 | 0.24 | 81 | 20.82 | 3 | 2 | 0.50 |
| C109.25.30 | 16 | 821.1[c] | 821.3 | 0.024 | 0.30 | 81 | 20.82 | 3 | 2 | 0.50 |
| C201.25.30 | 16 | 909.8[b] | 912.8 | 0.330 | 2.01 | 81 | 20.82 | 3 | 2 | 0.50 |
| C202.25.30 | 16 | 909.8[b] | 912.8 | 0.330 | 0.17 | 81 | 20.82 | 3 | 2 | 0.50 |
| C203.25.30 | 16 | 909.8[c] | 912.8 | 0.330 | 0.44 | 81 | 20.82 | 3 | 2 | 0.50 |
| C204.25.30 | 16 | 909.8[c] | 912.8 | 0.330 | 0.20 | 81 | 20.82 | 3 | 2 | 0.50 |
| C205.25.30 | 16 | 909.8[b] | 912.8 | 0.330 | 0.38 | 81 | 20.82 | 3 | 2 | 0.50 |
| C206.25.30 | 16 | 909.8[b] | 912.8 | 0.330 | 0.17 | 81 | 20.82 | 3 | 2 | 0.50 |
| C207.25.30 | 16 | 909.8[c] | 912.8 | 0.330 | 5.65 | 109 | 28.02 | 3 | 3 | 0.50 |
| C208.25.30 | 16 | 909.8[c] | 912.8 | 0.330 | 0.62 | 81 | 20.82 | 3 | 2 | 0.50 |
| C101.50.30 | 29 | 1599.5[d] | 1599.6 | 0.006 | 9.27 | 147 | 17.67 | 4 | 2 | 0.50 |
| C102.50.30 | 29 | 1599.5[d] | 1599.6 | 0.006 | 1.41 | 147 | 17.67 | 4 | 2 | 0.50 |
| C103.50.30 | 29 | 1598.3[c] | 1598.4 | 0.006 | 2.66 | 147 | 17.67 | 4 | 2 | 0.50 |
| C104.50.30 | 29 | 1598.3[c] | 1598.4 | 0.006 | 1.09 | 147 | 17.67 | 4 | 2 | 0.50 |
| C105.50.30 | 29 | 1599.5[d] | 1599.6 | 0.006 | 12.53 | 198 | 23.80 | 4 | 3 | 0.50 |
| C106.50.30 | 29 | 1599.5[d] | 1599.6 | 0.006 | 4.00 | 147 | 17.67 | 4 | 2 | 0.50 |
| C107.50.30 | 29 | 1599.5[d] | 1599.6 | 0.006 | 4.59 | 147 | 17.67 | 4 | 2 | 0.50 |
| C108.50.30 | 29 | 1598.3[d] | 1598.4 | 0.006 | 52.71 | 147 | 17.67 | 4 | 2 | 0.50 |
| C109.50.30 | 29 | 1598.3[c] | 1598.4 | 0.006 | 6.97 | 147 | 17.67 | 4 | 2 | 0.50 |
| RC101.25.30 | 18 | 1419.8[b] | **1419.8** | **0.000** | 4.87 | 102 | 21.98 | 5 | 2 | 0.50 |
| RC102.25.30 | 18 | 1419.8[b] | **1419.8** | **0.000** | 3.03 | 102 | 21.98 | 5 | 2 | 0.50 |
| RC103.25.30 | 18 | 1419.8[b] | **1419.8** | **0.000** | 3.00 | 102 | 21.98 | 5 | 2 | 0.50 |
| RC104.25.30 | 18 | 1419.8[b] | **1419.8** | **0.000** | 0.68 | 102 | 21.98 | 5 | 2 | 0.50 |
| RC105.25.30 | 18 | 1419.8[b] | **1419.8** | **0.000** | 4.86 | 141 | 30.39 | 5 | 3 | 0.50 |
| RC106.25.30 | 18 | 1419.8[b] | **1419.8** | **0.000** | 6.62 | 123 | 26.51 | 5 | 2 | 0.00 |
| RC107.25.30 | 18 | 1419.8[b] | **1419.8** | **0.000** | 1.60 | 102 | 21.98 | 5 | 2 | 0.50 |
| RC108.25.30 | 18 | 1419.8[b] | **1419.8** | **0.000** | 1.15 | 102 | 21.98 | 5 | 2 | 0.50 |
| RC201.25.30 | 18 | 1419.8[b] | **1419.8** | **0.000** | 13.64 | 141 | 30.39 | 5 | 3 | 0.50 |
| RC202.25.30 | 18 | 1419.8[b] | **1419.8** | **0.000** | 3.13 | 123 | 26.51 | 5 | 2 | 0.00 |
| RC203.25.30 | 18 | 1419.8[b] | **1419.8** | **0.000** | 5.16 | 102 | 21.98 | 5 | 2 | 0.50 |
| RC204.25.30 | 18 | 1419.8[b] | **1419.8** | **0.000** | 3788.68 | 102 | 21.98 | 5 | 2 | 0.50 |
| RC205.25.30 | 18 | 1419.8[b] | **1419.8** | **0.000** | 4.44 | 169 | 36.42 | 5 | 3 | 0.00 |
| RC206.25.30 | 18 | 1419.8[b] | **1419.8** | **0.000** | 7.73 | 102 | 21.98 | 5 | 2 | 0.50 |
| RC207.25.30 | 18 | 1419.8[b] | **1419.8** | **0.000** | 16.66 | 102 | 21.98 | 5 | 2 | 0.50 |
| RC208.25.30 | 18 | 1419.8[c] | **1419.8** | **0.000** | 2.80 | 141 | 30.39 | 5 | 3 | 0.50 |
| RC101.50.30 | 33 | 2739.5[c] | **2739.5** | **0.000** | 26.65 | 172 | 18.09 | 6 | 2 | 0.50 |
| RC102.50.30 | 33 | 2739.5[c] | **2739.5** | **0.000** | 16.68 | 172 | 18.09 | 6 | 2 | 0.50 |
| RC103.50.30 | 33 | 2739.5[c] | **2739.5** | **0.000** | 67.46 | 238 | 25.03 | 6 | 3 | 0.50 |
| RC104.50.30 | 33 | 2739.5[c] | **2739.5** | **0.000** | 110.02 | 172 | 18.09 | 6 | 2 | 0.50 |
| RC105.50.30 | 33 | 2739.6[c] | **2739.6** | **0.000** | 25.38 | 172 | 18.09 | 6 | 2 | 0.50 |
| RC106.50.30 | 33 | 2739.5[c] | **2739.5** | **0.000** | 42.26 | 172 | 18.09 | 6 | 2 | 0.50 |
| RC107.50.30 | 33 | 2739.5[c] | **2739.5** | **0.000** | 7.73 | 172 | 18.09 | 6 | 2 | 0.50 |
| RC108.50.30 | 33 | 2739.5[c] | **2739.5** | **0.000** | 3.95 | 172 | 18.09 | 6 | 2 | 0.50 |
| RC201.50.30 | 33 | 2739.5[c] | **2739.5** | **0.000** | 136.74 | 235 | 24.71 | 6 | 2 | 0.00 |
| RC202.50.30 | 33 | 2739.5[c] | **2739.5** | **0.000** | 18.04 | 172 | 18.09 | 6 | 2 | 0.50 |
| RC203.50.30 | 33 | 2739.5[c] | **2739.5** | **0.000** | 62.54 | 172 | 18.09 | 6 | 2 | 0.50 |
| RC204.50.30 | 33 | 2739.5[c] | **2739.5** | **0.000** | 15.94 | 172 | 18.09 | 6 | 2 | 0.50 |
| RC205.50.30 | 33 | 2739.5[c] | **2739.5** | **0.000** | 96.10 | 235 | 24.71 | 6 | 2 | 0.00 |
| RC206.50.30 | 33 | 2739.5[c] | **2739.5** | **0.000** | 27.41 | 235 | 24.71 | 6 | 2 | 0.00 |
| RC207.50.30 | 33 | 2739.5[c] | **2739.5** | **0.000** | 16.11 | 172 | 18.09 | 6 | 2 | 0.50 |
| RC208.50.30 | 33 | 2739.5[c] | **2739.5** | **0.000** | 3.00 | 172 | 18.09 | 6 | 2 | 0.50 |
| Average | | | | 0.084 | 81.71 | | 21.72 | 4.54 | 2.14 | 0.44 |

[a] Intel Xeon E5-2640 v3, 2.6 GHz
[b] Desaulniers et al. [20]
[c] Archetti et al. [7]
[d] Luo et al. [25]

Table 4: LKH-3 results for SDVRPTW with limited fleet on the Group C and RC instances of Solomon [33]

An interesting future direction of this work would be to consider the full Pareto frontier defined by the trade-off between the size of a partition $\mu$ and the family of partitions that it coalesces to. That is, while we required in this paper that $\mu$ be lossless, it might also be sensible to use partitions that are not entirely lossless, but very close to being so, if their cardinality is sufficiently small.

## 6 Acknowledgement

The authors acknowledge the Center for Advanced Research Computing (CARC) at the University of Southern California for providing computing resources that have contributed to the research results reported within this publication. URL: https://carc.usc.edu.